\documentclass{ws-ijmpa}
\def\oo{\infty}

\def\SYS{{\tt SYS}}

\def\Cl{{\mathrm{Cl}}}

\def\PL{{\it Phys. Lett. }}

\def\IJMP{{\it Int. J. Mod. Phys. }}

\def\dq#1{d^Dq_{#1}}
\def\Dq#1{[d^Dq_{#1}]}
\def\e{\epsilon}

\def\Gammaq{\Gamma\left(\dfrac{1}{15}\right)\Gamma\left(\dfrac{2}{15}\right)\Gamma\left(\dfrac{4}{15}\right)\Gamma\left(\dfrac{8}{15}\right)}
\def\XOMOG{X^{HO}}
\def\XNOMG{X^{NH}}
\def\ru{\sqrt{u}}

\def\Fcub#1{{}_2F_1\left(\begin{smallmatrix}
{{ \frac{1}{3}\;\frac{2}{3} }}\\
{{1 }}\end{smallmatrix}; #1\right)}
\def\Ecub#1{{}_2F_1\left(\begin{smallmatrix}
{{ \frac{1}{3}\;-\frac{1}{3} }}\\
{{1 }}\end{smallmatrix}; #1\right)}
\def\Fqt#1{{}_4F_3\left(\begin{smallmatrix}
{{ \frac{1}{2}\;\frac{1}{2}\;\frac{1}{2}\;\frac{1}{2} }}\\
{{1\;1\;1 }}\end{smallmatrix}; #1\right)}
\def\Fqq#1{{}_2F_1^2\left(\begin{smallmatrix}
{{ \frac{1}{4}\;\frac{1}{4} }}\\
{{1 }}\end{smallmatrix}; #1\right)}
\def\TITLE{ANALYTICAL EXPRESSIONS OF 3 AND 4-LOOP SUNRISE FEYNMAN INTEGRALS AND
4-DIMENSIONAL LATTICE INTEGRALS}
\def\CAPTIONFIGA{Three-loop and four-loop self-mass diagrams.}
\def\eqref#1{Eq.(\ref{#1})}
\def\eqrefb#1#2{Eqs.(\ref{#1})-(\ref{#2})}
\def\itref#1{(\ref{#1})}
\def\itrefb#1#2{(\ref{#1})-(\ref{#2})}
\def\FigureOne
{
\begin{figure}
\begin{center}
   \begin{picture}(150,130)(0,0)
   \thicklines
   \put(075,010){\makebox(0,0)[lb]{
   \put(-040,000){\line(-1,0){15}}
   \put(+040,000){\line(+1,0){15}}
   \qbezier(+000.0,+040.0)(-022.7,+039.3)(-034.6,+020.0)
   \qbezier(-034.6,+020.0)(-045.4,+000.0)(-034.6,-020.0)
   \qbezier(-034.6,-020.0)(-022.7,-039.3)(+000.0,-040.0)
   \qbezier(+000.0,-040.0)(+022.7,-039.3)(+034.6,-020.0)
   \qbezier(+034.6,-020.0)(+045.4,+000.0)(+034.6,+020.0)
   \qbezier(+034.6,+020.0)(+022.7,+039.3)(+000.0,+040.0)
\qbezier(-040.0,+000.0)(+000.0,+040.0)(+040.0,+000.0)
   \qbezier(-040.0,+000.0)(+000.0,-040.0)(+040.0,+000.0)
  \put(-066,-005){\text{$p$}}   
  \put(-007,+033){\text{$m_1$}}   
  \put(-007,+012){\text{$m_2$}}   
  \put(-007,-016){\text{$m_3$}}   
  \put(-007,-036){\text{$m_4$}}   
  \put(-012,-070){\text{\LARGE$S_3$}}   
   }}
   \end{picture}
   \begin{picture}(150,130)(0,0)
   \thicklines
   \put(075,010){\makebox(0,0)[lb]{
   \put(-040,000){\line(-1,0){15}}
   \put(+040,000){\line(+1,0){15}}
   \qbezier(+000.0,+040.0)(-022.7,+039.3)(-034.6,+020.0)
   \qbezier(-034.6,+020.0)(-045.4,+000.0)(-034.6,-020.0)
   \qbezier(-034.6,-020.0)(-022.7,-039.3)(+000.0,-040.0)
   \qbezier(+000.0,-040.0)(+022.7,-039.3)(+034.6,-020.0)
   \qbezier(+034.6,-020.0)(+045.4,+000.0)(+034.6,+020.0)
   \qbezier(+034.6,+020.0)(+022.7,+039.3)(+000.0,+040.0)
 \qbezier(-040.0,+000.0)(+000.0,+000.0)(+040.0,+000.0)
 \qbezier(-040.0,+000.0)(+000.0,+050.0)(+040.0,+000.0)
 \qbezier(-040.0,+000.0)(+000.0,-050.0)(+040.0,+000.0)
  \put(-066,-005){\text{$p$}}   
  \put(-007,+033){\text{$m_1$}}   
  \put(-007,+017){\text{$m_2$}}   
  \put(-007,-007){\text{$m_3$}}   
  \put(-007,-020){\text{$m_4$}}   
  \put(-007,-036){\text{$m_5$}}   
  \put(-012,-070){\text{\LARGE$S_4$}}   
   }}
   \end{picture}
\end{center}
\phantom{ }\vspace{-1.1truecm}\phantom{ } 
\caption{\CAPTIONFIGA}
 \label{figureone}
 \end{figure}
}
\newcommand\mytoday{\number\day\space \ifcase\month\or
  January\or February\or March\or April\or May\or June\or
    July\or August\or September\or October\or November\or December\fi
      \space\number\year}
\begin{document}
 \markboth{S. Laporta}{\TITLE}                        

\catchline{}{}{}{}{} 

\title{\TITLE }
\author{S. Laporta}
\address{Museo Storico della Fisica e Centro Studi e Ricerche Enrico Fermi,
\\ 
Dipartimento di Fisica, Universit\`a di Bologna,\\
INFN, Sezione di Bologna,\\
Via Irnerio 46, I-40126 Bologna, Italy \\
laporta@bo.infn.it}
\maketitle

\begin{history} 
\received{Day Month Year} 
\revised{Day Month Year} 
\end{history} 

\begin{abstract}
In this paper we continue the work begun in 2002 on the
identification of the analytical expressions of Feynman integrals
which require the evaluation of multiple elliptic integrals.
We rewrite and simplify the analytical expression of the 3-loop  
self-mass integral with three equal masses and on-shell external momentum.
We collect and analyze a number of results on double and triple elliptic
integrals. By using very high-precision numerical fits, for the first time
we are able to identify a
very compact analytical expression for the 4-loop on-shell self-mass integral
with 4 equal masses,
that is one of the master integrals of the 4-loop electron $g$-$2$.
Moreover, we fit the analytical expressions of some integrals which
appear in lattice perturbation theory, and in particular the 4-dimensional
generalized Watson integral.
\end{abstract}


\ccode{PACS numbers: 02.60.Jh, 12.20Ds, 12.38.Gc}

\pagenumbering{roman}
\setcounter{page}{0}
\vfill\eject 
\pagenumbering{arabic}
\setcounter{page}{1}
\section{Introduction}
Analytical expressions of many Feynman diagrams 
contain polylogarithmic functions of various kinds 
(Nielsen polylogarithms,
Harmonic polylogarithms\cite{Remiddi:1999ew},
Harmonic sums\cite{Vermaseren:1998uu,Blumlein:1998if},
multiple zeta values, Euler sums\cite{Broadhurst:1996kc}, etc{\ldots}). 
But there exist Feynman integrals which
cannot be described only in terms of polylogarithms.

At two-loop level, the discontinuity of the off-shell massive 
``sunrise'' diagram with different masses is expressed by elliptic
functions. At three (or more) loop level the situation worsens.
These diagrams contain nested multiple elliptic integrals;
the current mathematical knowledge of such integrals is scarce or missing.
At this preliminary stage, \emph{``experimental mathematics''}
is the best tool. In other words, high-precision numerical values of the 
integrals are fitted\cite{Lapotv4} 
with various candidate analytical expressions 
until an agreement is found. The equalities are then checked up to hundreds 
or thousands of digits.
In this way the right analytical expression is identified beyond any reasonable
doubt. This may be the starting point of the subsequent search for 
a rigorous proof of the result, task which may take months 
of hard work\cite{bro1,bro2,talkbroadhurst2}.

In Ref.~\refcite{Lapotv4} we work out a very high-precision value of the 
3-loop scalar master integral of the ``sunrise'' diagram $S_3$ of Fig.1; then we 
fit that value with products of elliptic integrals,
checking the equality with a precision of thousand of digits. 

In this paper we continue the work on the 3-loop integral, 
and we simplify the 3-loop analytic result by using some identities
between elliptic integrals.
Next we review and develop the approach used for fitting and
identifying the 3-loop integral and we apply it to the 
4-loop scalar master integral of the ``sunrise'' diagram $S_4$ of Fig.1.
The analytical calculation of this master integral is also of physical interest, 
because it is the simplest non-trivial master integral of the
4-loop electron $g$-$2$ (of which a high-precision numerical calculation
is under way\cite{Lapotalk}).
Very likely, the analytical constants which appear in the expression of $S_4$
should also appear in the analytical expression of 4-loop $g$-$2$.
We calculate an high-precision value of $S_4$ and we are able
to fit this value with an expression containing two new elliptic constants,
checking the equality with a precision of thousand of digits. 

We apply this procedure also to the values of some 4-dimensional
lattice integrals, and we identify their analytical expressions;
surprisingly, they contain the same elliptic
constants of the 4-loop integral.

The plan of the paper is the following:
In section \ref{treloop} we simplify the results of
Ref.~\refcite{Lapotv4} by using identities between elliptic integrals.
In section \ref{quattroloop} we study the 4-loop ``sunrise'' integral.
We collect a number of results on a ``simplified'' version of the integrals
involved. Then we use these results as a guide for identifying the 
candidate analytical expressions suitable for fitting the 4-loop results.
In section \ref{risultati} we show the analytical results found for the 4-loop
integrals. In section \ref{quattrolattice} we fit the values of
some 4-dimensional lattice integrals with the same analytical constants 
discovered in  the 4-loop integrals.
In section \ref{conclusions} we give our conclusions.
\FigureOne
\section{Three-loop single-scale self-mass integral}\label{treloop}
\subsection[subtitle]{The results of Ref.~\refcite{Lapotv4}}
In Ref.~\refcite{Lapotv4} we considered the Feynman diagram 
$S_3(p^2,m_1^2,m_2^2,m_3^2,m_4^2,D)$ with equal masses $m_j=1$, and on-shell
external momentum (see Fig.1)
\begin{multline}
S_3(-1,1,1,1,1,D)=
\int\dfrac{\Dq1\;\Dq2\;\Dq3}{(q_1^2+1)(q_2^2+1)(q_3^2+1)((p-q_1-q_2-q_3)^2+1)}\ ,
\\\quad p^2=-1\ ,
\end{multline} 
where 
\begin{equation}
\Dq{}= \dfrac{\dq{}}{\pi^{D/2} \Gamma\left(3-\dfrac{D}{2}\right) }\ .
\end{equation} 
By using an hyperspherical representation for the integral, 
we found that 
the 
value of $S_3$ could be expressed as a sum of various
double elliptic integrals,
the simplest being
\begin{multline}\label{EllA} 
A_3=\int^\oo_0 \dfrac{dl}{R(l,-1,-1)} 
\int^\oo_0 \dfrac{dm}{ R(m,l,-1) R
(m,-1,-1)}=
\\2.641\;379\;476\;074\;689\;431\;349{\ldots} \ , 
\end{multline} 
\begin{equation} 
R(x,y,z)=\sqrt{x^2+y^2+z^2-2xy-2xz-2yz}\ .
\end{equation} 
We were not able to calculate $A_3$ in analytical form. 
Therefore we evaluated it at very high precision and we tried to
fit the numerical value with various kinds of analytical expressions.
In Ref.~\refcite{Lapotv4} we found that 
\begin{equation}\label{A3}
A_3=K(w_{-})K(w_{+})\ , 
\qquad w_{\pm}=\dfrac{z_{\pm}}{z_{\pm}-1}\ ,
\qquad z_{\pm}=-(2-\sqrt{3})^4(4\pm\sqrt{15})^2\ ,
\end{equation}
where $K$ is the first of the two elliptic integrals 
\begin{equation}
K(m)=\int^1_0 \dfrac{dt}{\sqrt{1-t^2}\sqrt{1-m t^2}}\ , 
\qquad 
E(m)=\int^1_0 \dfrac{dt\;\sqrt{1-m t^2}}{\sqrt{1-t^2}}\ .
\end{equation}
We were able to fit the values of
$S_3$ in 2 and 4 dimensions:\footnote{ 
The two-dimensional value \itref{S3_due} was calculated together with
\itref{S3_quattro}, but not published in Ref.~\refcite{Lapotv4}.}
\begin{equation}\label{S3_due}
S_3(-1,1,1,1,1,D=2)=\frac{4\pi}{\sqrt{15}}K(w_{-})K(w_{+})\ ,
\end{equation}
\begin{multline}\label{S3_quattro}
S_3(-1,1,1,1,1,D=4-2\epsilon)=
 2\e^{-3}+\dfrac{22}{3}\e^{-2} + \dfrac{577}{36}\e^{-1}
+\dfrac{4\pi}{\sqrt{15}}
  \left(
 \dfrac{35}{8}\pi 
  \right.\\\left.
 +\dfrac{131}{12}K(w_{-})K(w_{+})
 -\dfrac{7}{2}
 \left(E(1-w_{-})E(1-w_{+})+5E(w_{-})E(w_{+})\right)
 \right)
 +\dfrac{6191}{216} +O(\e) . \\
\end{multline} 
\eqref{A3},
\eqref{S3_due} and \eqref{S3_quattro} were checked up to 30000, 40000 and 1200 digits,
respectively. 
\subsection{New relations between elliptic integrals}
Now we note that the arguments $w_{\pm}$ 
of the elliptic integrals are \emph{singular values}. 
In the context of elliptic integrals
$k_r$ is a called singular value if 
\begin{equation}
 \dfrac{K(1-k_r)}{K(k_r)}= \sqrt{r} \ ,
\end{equation}
where $r$ is an integer or a rational number.
The arguments $w_{\pm}$ of elliptic integrals are
singular values for $r=15$ and $r=5/3$ respectively,
\begin{equation*}
w_{-}=k_{15}\ , \qquad w_{+}=k_{5/3}\ ,
\end{equation*} 
that is 
\begin{equation}\label{end1}
 \dfrac{K(1-k_{15})}{K(k_{15})}=\sqrt{15}\ , \quad
 \dfrac{K(1-k_{5/3})}{K(k_{5/3})}=\sqrt{\dfrac{5}{3}} \ ,
\end{equation} 
\begin{equation}\label{a1553}
 \dfrac{K(k_{5/3})}{K(k_{15})}= \dfrac{\sqrt{15}-\sqrt{3}}{2}\ . 
\end{equation} 
The values of elliptic integrals of second kind 
of \eqref{S3_quattro} are obtained following Ref.~\refcite{Borwein}
\begin{align}
E(k_r)  &=\dfrac{\pi}{4\sqrt{r}K(k_r)}
+K(k_r)\left(1-\dfrac{\alpha_r}{\sqrt{r}}\right) \ ,\\
E(1-k_r)&=\dfrac{\pi}{4K(k_r)}+K(k_r)\alpha_r \ .\\
\end{align}
By using the values~\cite{Borwein}
\begin{equation}\label{end2}
\alpha_{15} =\dfrac{\sqrt{15}-\sqrt{5}-1}{2}\ ,\quad
\alpha_{5/3}=\dfrac{\sqrt{15}-\sqrt{5}+1}{6}\ ,
\end{equation}
\begin{equation}
K(k_{15})=\sqrt{ \dfrac{(\sqrt{5}+1)P}{240\pi}}, 
\end{equation}
where 
\begin{equation}\label{quattrogamma}
P\equiv \Gammaq\ ,
\end{equation}
and expressing $K(k_{5/3})$ by using \eqref{a1553},
we are able to rewrite \eqref{S3_due} and \eqref{S3_quattro} in the very
compact form  
\begin{equation}\label{S3_due_res}
S_3(-1,1,1,1,1,D=2)=\dfrac{P}{40\sqrt{3}\pi}\ ,
\end{equation}
\begin{equation}\label{S3_quattro_res}
S_3(-1,1,1,1,1,D=4-2\epsilon)=
 2\e^{-3}+\dfrac{22}{3}\e^{-2} + \dfrac{577}{36}\e^{-1}
 +\dfrac{6191}{216}  -\dfrac{14\sqrt{5}\pi^4}{P}
 -\dfrac{\sqrt{5}}{900}{P}
 +O(\e) .
\end{equation}
\eqrefb{end1}{end2} were also shown by David~Broadhurst in his  beautiful talk given 
in Bielefeld\cite{talkbroadhurst}.
After the talk,
our unpublished results \itrefb{quattrogamma}{S3_quattro_res} were shown to
David~Broadhurst.
\subsection{The path to \eqref{A3}}\label{trickA3}
For sake of completeness
we recall here some unpublished observations that suggested us the form of \eqref{A3}.
First we calculated analytically the \emph{simplest} double elliptic integral:
\begin{equation}\label{A3bis}
   \int^1_0 \dfrac{dx}{\sqrt{1-x^2}}
   \int^1_0 \dfrac{dy}{\sqrt{1-y^2}\sqrt{1-x^2 y^2}}=
   \left[K\left(\dfrac{1}{2}\right)\right]^2= 3.437\;592\;909{\ldots} \ ,
\end{equation} 
which factorizes in the square of the elliptic integral $K$.

Then we observed that in the diagram $S_3(-1,1,1,1,1)$ the value of the
square of the external momentum $p^2$ is $-1$, which
is different from the threshold ($p^2=-16$) or the 
pseudothresholds ($p^2=-4,0$). 
We expected that the analytical expression is much simpler 
for the on-threshold diagram 
than for the off-threshold diagram.
So we considered the above graph with one mass changed: $S_3(-1,1,1,1,2)$.
Now the value of $p^2=-1$ is on a pseudothreshold  (which are
at $p^2=-1,-9,-25$).
The integral analogous to $A_3$ 
is 
\begin{equation}\label{EllA_mod} 
{A'}_3=\int^\oo_0 \dfrac{dl}{R(l,-1,-1)} 
\int^\oo_0 \dfrac{dm}{ R(m,l,-4) R (m,-1,-1)}=1.474\;585\;992\;{\ldots} \ . 
\end{equation} 
We were able to fit the numerical value of ${A'}_3$  with 
\begin{equation}\label{A3_mod} 
{A'}_3
=\frac{1}{\sqrt{3}}\left(\frac{\pi}{2}\right)^2 \Fqq{\frac{1}{4}}
=\frac{1}{\sqrt{3}}\left[K\left(\frac{2-\sqrt{3}}{4}\right)\right]^2
=\dfrac{\Gamma^6\left(\frac{1}{3}\right)}{2^\frac{14}{3}\pi^2}\ .
\end{equation}
Subsequently, as we expected the form of $A_3$ to be more complicated 
than ${A'}_3$,
we tried also products of $K$ with different arguments, and we found \eqref{A3}.
\section{Four-loop single-scale self-mass integral}\label{quattroloop}
Now we consider the single-scale 4-loop self-mass diagram
$S_4(p^2,m_1^2,m_2^2,m_3^2,m_4^2,D)$ of Fig.1,
in the case of equal masses $m_j=1$ and on shell external momentum
$p^2=-1$. This diagram has 3 master integrals, the simplest being
\begin{multline}\label{S4}
S_4(-1,1,1,1,1,1,D)=\\
\int\dfrac{\Dq1\;\Dq2\;\Dq3\;\Dq4}
{(q_1^2+1)(q_2^2+1)(q_3^2+1)(q_4^2+1)((p-q_1-q_2-q_3-q_4)^2+1)}\ .
\end{multline} 
As already said in the introduction, 
this is also one of the several master integrals which appear 
in the calculation of 4-loop $g$-$2$.

The first observation is that, fortunately, the value of $p^2=-1$ in $S_4$ is already 
on a pseudo\-threshold (which are $p^2=-1$,$-9$,$-25$);
therefore to simplify the analytical structure of the integral
we have not to use the mass change of the previous section \ref{trickA3}. 

\subsection{High-precision numerical values}
We need an high-precision numerical value of this integral
in order to obtain a meaningful fit.
This is worked out by using the difference equation method presented 
in Refs.~\refcite{Lapdif1,Lapdif2}. 
Summarizing, one raises to $n$ one denominator of \eqref{S4}
\begin{equation}
X_4(n)=
\int\dfrac{\Dq1\;\Dq2\;\Dq3\;\Dq4}{(q_1^2+1)^n(q_2^2+1)(q_3^2+1)(q_4^2+1)((p-q_1-q_2-q_3-q_4)^2+1)}\\
\ .
\end{equation}
The function $X_4(n)$ satisfies the fourth-order difference equation 
\begin{multline}\label{eqdif}
p_1 X_4(n+3) +p_2 X_4(n+2) +p_3 X_4(n+1) +p_4 X_4(n) +p_5 X_4(n-1) =\\
24(D-2)^4 J^3(1)J(n) \ ,
\end{multline}
where
\begin{equation}
\begin{split}
p_1&=- 768 n (n+1) (n+2) (n-2D+5) \ ,\\
p_2&= 128n(n+1) \bigl(  11 n^2 +( 63 - 26 D )n  + 11 D^2  - 57 D + 76 \bigr) \ ,\\
p_3&=4n\bigl(  - 129 n^3  + (294 D - 588)n^2 
         +( - 148 D^2 + 592 D - 567   )n  \\
	& + 8 D^3  - 48 D^2  + 46 D   + 36  \bigr)\ , \\
p_4&=2\bigl( - 60n^4   +(294 D  - 468 ) n^3  +(- 485 D^2 + 1499 D - 1164 )n^2 
     +(  308 D^3 \\& - 1363 D^2 + 2015 D - 996 )n  - 60 D^4  + 326 D^3 - 658 D^2 + 584 D - 192 \bigr) \ , \\ 
p_5&=- (n-D+1)(n-2D+3)(2n-3D+4)(2n-5D+8)\ , \\
\end{split}
\end{equation}
and $J$ is the one-loop integral
\begin{equation}\label{defj}
J(n)= \int\dfrac{\Dq1}{(q_1^2+1)^n} \ .
\end{equation}
\eqref{eqdif} contains in the r.h.s. the integral obtained from $S_4$ by
contracting one
line, which factorizes into the product of 4 one-loop tadpoles. The solution
of \eqref{eqdif} compatible with the large-$n$ boundary condition
$X_4(n)\propto n^{-D/2}$ can be written as $ X_4(n)=C_1 \XOMOG_1(n) +C_2
\XOMOG_2(n) +\XNOMG(n) $.
The functions $\XOMOG_1$, $\XOMOG_2$ and $\XNOMG$ are respectively the two solutions of the
homogeneous equation compatible with
the above large-$n$ behaviour and one particular solution of the nonhomogeneous
equation \eqref{eqdif}. The constants $C_1$ and $C_2$ are obtained from 
the 3-loop self-mass integrals belonging to the diagram obtained from $S_4$ 
by deleting one line, that is $S_3$. 
The amount of calculations needed to work out and solve the systems of 
difference equations is high, so that the calculations
have been performed by means of an automatic tool, the program {\SYS } 
described in Ref.~\refcite{Lapdif1}.                                           

In two dimension one finds 
\begin{equation} \label{S4_due}
S_4(-1,1,1,1,1,1,D=2)= 40.2451219019305821264798187417 \ldots \ 
\end{equation}
and  in the limit $D\to 4$
\begin{align}\label{S4_quattro}
S_4&(-1,1,1,1,1,1,D=4-2\epsilon)= 
-\frac{5}{2\epsilon^4} 
-\frac{45}{4\epsilon^3} 
-\frac{4255}{144\epsilon^2} 
-\frac{106147}{1728\epsilon} \\
   &-141.72215618664768694996791  
    -521.14654568600250441775466\epsilon  \nonumber \\
  &-3347.9933650782886117865341\epsilon^2 
  -17951.3774774809944931097622\epsilon^3 \nonumber\\
&-101753.8165331173182139560386\epsilon^4 +O(\epsilon^5)\ .
\end{align}
All the above numerical constants were calculated with a precision of over 2400
digits; for the sake of space we show here only the first 25.

\subsection{Triple elliptic integrals}
By using an hyperspherical representation for the integral, 
$S_4(-1,1,1,1,1,1,D=2)$ and the finite part of $S_4(-1,1,1,1,1,1,D=4-2\e)$
contain \emph{triple} elliptic integrals,
the simplest being
\begin{multline}\label{A4}
A_4=\int^\oo_0 \dfrac{dl}{R(l,-1,-1)} 
\int^\oo_0  \dfrac{dm}{ R(m,l,-1) }
\int^\oo_0\dfrac{dr}{R(m,r,-1)R(r,-1,-1)}=\\8.749\;361\;490{\ldots}  \ .
\end{multline} 
We prefer to redistribute the arguments of the $R$ functions and 
consider the similar integral  
\begin{multline}\label{A4bis}
\int^\oo_0 \dfrac{dl}{R(l,-1,-1)} 
\int^\oo_0  \dfrac{dm}{ R(m,-1,-1) }
\int^{\left(\sqrt{l}-\sqrt{m}\right)^2}_0\dfrac{dr}{R(r,l,m)R(r,-1,-1)}=\\ i
8.749\;361\;490{\ldots}= i A_4 \ ,
\end{multline} 
and the companion integral
\begin{multline}\label{B4}
B_4=\int^\oo_0 \dfrac{dl}{R(l,-1,-1)} 
\int^\oo_0  \dfrac{dm}{ R(m,-1,-1) }
\int^{\left(\sqrt{l}+\sqrt{m}\right)^2}_{\left(\sqrt{l}-\sqrt{m}\right)^2}
\dfrac{dr}{R(r,l,m)R(r,-1,-1)}=\\
9.607\;815\;129{\ldots}\ ,
\end{multline} 
where $r_{\pm}=\left(\sqrt{l}\pm\sqrt{m}\right)^2$ are the two zeroes of $R(r,l,m)$.
$A_4$ and $B_4$ are the 4-loop integrals analogues of the 3-loop integral $A_3$,
and  we have to find their analytical expressions.
\subsection{Simplifying the problem}
First of all we consider the simplest triple elliptic integral with
structure similar to \eqref{A4bis}: 
\begin{equation}\label{Adef}
A= \int^1_0 \dfrac{dx}{\sqrt{1-x^2}}
   \int^1_0 \dfrac{dy}{\sqrt{1-y^2}}
   \int^1_0 \dfrac{dz}{\sqrt{1-z^2} \sqrt{1-x^2 y^2
   z^2}}=4.335\;593\;665{\ldots} \ .
\end{equation} 
By changing one limit of integration over $z$ to the zero of $1-x^2 y^2 z^2$
we obtain a companion integral analogous to \eqref{B4}
\begin{equation}\label{Bdef}
B= \int^1_0 \dfrac{dx}{\sqrt{1-x^2}}
   \int^1_0 \dfrac{dy}{\sqrt{1-y^2}}
   \int^{1/xy}_1 \dfrac{dz}{\sqrt{z^2-1} \sqrt{1-x^2 y^2
   z^2}}=6.997\;563\;016{\ldots} \ .
\end{equation} 
We expect that the study of the simpler constants
$A$ and $B$ 
can help us understand the analytical expressions of $A_4$ and $B_4$.
Integrating over $z$ 
\begin{equation}
A= \int^1_0 \dfrac{dx}{\sqrt{1-x^2}}
   \int^1_0 \dfrac{dy}{\sqrt{1-y^2}} K(x^2 y^2) \ ,
\end{equation} 
\begin{equation}
B= \int^1_0 \dfrac{dx}{\sqrt{1-x^2}}
   \int^1_0 \dfrac{dy}{\sqrt{1-y^2}} K(1-x^2 y^2)\ .
\end{equation} 
Integrating over $y$ and $x$
\begin{equation}
A
=
\int^1_0 \dfrac{dx}{\sqrt{1-x^2}} 
K^2\left(\dfrac{1-\sqrt{1-x^2}}{2}\right)
=
\left(\dfrac{\pi}{2}\right)^3   \Fqt{1}\ ,
\end{equation} 

\begin{equation}
B= 
\int^1_0 \dfrac{dx}{\sqrt{1-x^2}} 
K\left(\dfrac{1-\sqrt{1-x^2}}{2}\right)
K\left(\dfrac{1+\sqrt{1-x^2}}{2}\right)\ .
\end{equation} 
No analytical expression is known for $\Fqt{1}$.
Trying to understand the reason of that, 
we study the following family of integrals:

\begin{equation}\label{family}
\int^1_0 dt \; K^m(t) \; K^n(1-t)\;
\left({\dfrac{1}{\sqrt{t}}}\right)^\alpha
\left({\dfrac{1}{\sqrt{1-t}}}\right)^\beta \ .
\end{equation}
We consider here only the integrals which have results containing elliptic
constants. 
At the value $m+n=1$ there is  the integral 
\begin{equation}\label{I111}
\int^1_0 dt\;\dfrac{K(t)}{\sqrt{t(1-t)}}=2K^2\left(\dfrac{1}{2}\right) \ ,
\end{equation}
equivalent to \eqref{A3bis};  note  that it factorizes in a square of 
$K(1/2)=\Gamma^2(1/4)/(4\sqrt{\pi})$.

At the value $m+n=2$ we find numerically that the six integrals
\begin{equation}\label{I2001}
\int^1_0 dt\;\dfrac{K^2(t)}{\sqrt{t}}=B \ ,
\end{equation}
\begin{equation}\label{I2010}
\int^1_0 dt\;\dfrac{K^2(t)}{\sqrt{1-t}}=2B \ ,
\end{equation}
\begin{equation}\label{I2011}
\int^1_0 dt\;\dfrac{K^2(t)}{\sqrt{t(1-t)}}=4A \ ,
\end{equation}
\begin{equation}\label{I1101}
\int^1_0 dt\;\dfrac{K(t)K(1-t)}{\sqrt{t}}=\int^1_0 dt\;\dfrac{K(t)K(1-t)}{\sqrt{1-t}}=2A \ ,
\end{equation}
\begin{equation}\label{I1111}
\int^1_0 dt\;\dfrac{K(t)K(1-t)}{\sqrt{t(1-t)}}=2B \ ,
\end{equation}
can be expressed in terms of $A$ and $B$.
At the value $m+n=3$ we find a surprise:
six integrals factorizes into the fourth power of  $K(1/2)$.
\begin{equation}\label{I3000}
\int^1_0 dt\;K^3(t)=\dfrac{4}{5}K^4\left(\dfrac{1}{2}\right) \ ,
\end{equation}
\begin{equation}\label{3001}
\int^1_0
dt\;\dfrac{K^3(t)}{\sqrt{t}}=\dfrac{6}{5}K^4\left(\dfrac{1}{2}\right) \ ,
\end{equation}
\begin{equation}\label{3010}
\int^1_0 dt\;\dfrac{K^3(t)}{\sqrt{1-t}}=4 K^4\left(\dfrac{1}{2}\right) \ ,
\end{equation}
\begin{equation}\label{2100}
\int^1_0 dt\;K^2(t)K(1-t)=\dfrac{2}{3}K^4\left(\dfrac{1}{2}\right) \ ,
\end{equation}
\begin{equation}\label{2101}
\int^1_0
dt\;\dfrac{K^2(t)K(1-t)}{\sqrt{t}}=\dfrac{4}{3}K^4\left(\dfrac{1}{2}\right) \ ,
\end{equation}
\begin{equation}\label{2110}
\int^1_0 dt\;\dfrac{K^2(t)K(1-t)}{\sqrt{1-t}}=2 K^4\left(\dfrac{1}{2}\right)\ .
\end{equation}
The analytical results \itrefb{I2001}{2110}  have been fitted numerically and checked
up to 200 digits of precision.
We find results factorized at level 2 and 4, but not at
level 3 (odd).
This behaviour reminds us of the non-factorization of values of Riemann $\zeta$-function
at odd integers, and 
suggest to consider the constants $A$ and $B$ as irreducible objects.
We can relate $A$ and $B$ to multidimensional $\zeta$-like quadruple series.
Let us consider the integrals 
\begin{equation}
I_m=\int^1_0
dt\;\dfrac{K^2(t)}{\sqrt{1-t}}\left(\dfrac{K(1-t)}{K(t)}\right)^m \ .
\end{equation}
These integrals correspond to the integrals
\itref{I2010}, \itref{I1101}, \itref{I2001},  for $m=0$, $1$, $2$,  respectively,
and their values are  $I_0=2B$, $I_1=2A$, $I_2=B$.
We apply the change of variable 
\begin{equation}
q=\exp(-\pi K(1-t)/K(t)) \quad\text{or\ equivalently}\quad
1-t=(\theta_4(q)/\theta_3(q))^4\ ,
\end{equation}
where $\theta_j(q)$ are the Jacobi Theta Functions,
then
\begin{equation}
I_m= \pi^{2-m}
\int^1_0 dq  \left(
\theta_4^2(q)\theta_3(q)\dfrac{d}{dq}\theta_3(q)
-
\theta_3^2(q)\theta_4(q)\dfrac{d}{dq}\theta_4(q)
\right)(-\log q)^m\ .
\end{equation}
Expanding in series the $\theta$ functions, and integrating over $q$ 
term-by-term, $I_m$ becomes a quadruple series
\begin{equation}
I_m=m!\pi^{2-m} 
\sum_{i=-\oo}^{\oo}\sum_{j=-\oo}^{\oo}\sum_{k=-\oo}^{\oo}\sum_{l=-\oo}^{\oo}{}^{'}
\dfrac{(-1)^{i+j}(k^2-i^2)}{(i^2+j^2+k^2+l^2)^{m+1}}\ ;
\end{equation}
for $m=2$ the series converges, so that 
\begin{equation}
B=2
\sum_{i=-\oo}^{\oo}\sum_{j=-\oo}^{\oo}\sum_{k=-\oo}^{\oo}\sum_{l=-\oo}^{\oo}{}^{'}
\dfrac{(-1)^{i+j}(k^2-i^2)}{(i^2+j^2+k^2+l^2)^3}\ ,
\end{equation}
where the prime means that the origin $i=j=k=l=0$ must be excluded in the summation.
\subsection{Integrals of products of homogeneous solutions}

Coming back to the  4-loop integral,
if we close the 4-loop self-mass diagram $S_4$ by connecting together the
two external lines, we obtain a 5-loop vacuum diagram.
The 5-loop vacuum diagram can be decomposed into two 2-loop self-mass diagrams
connected together.
Therefore in $D=2$ dimensions
its value is given by the integral of the square of the 2-loop 
self-mass diagram $\int d^2 p \;S_2^2(p^2,1,1,1)$. 
The vacuum diagram is expected to have the same analytical 
structure of $S_4$, but with higher transcendentality.
$S_2(u)$ satisfies a second order nonhomogeneous differential equation
(see Ref.~\refcite{twoloopsunrise} for more details).
The corresponding homogeneous differential equation has two solutions
$J_1(u)$ and $J_2(u)$.
In order to reduce the transcendentality we may substitute 
$S_2(u)$ with $J_1(u)$ or $J_2(u)$.
The analytical expressions of $J_1$ and $J_2$ are given in the Appendix of 
Ref.~\refcite{twoloopsunrise};
for example, if $0 \le u \le 1$ they read
\begin{eqnarray}\label{exp01}
  J_1(u) &=& \frac{1}{\sqrt{(1+\ru)^3(3-\ru)}} K\bigl(a(u)\bigr) 
                                                  \ , \nonumber\\ 
  J_2(u) &=& \frac{1}{\sqrt{(1+\ru)^3(3-\ru)}} K\bigl(1-a(u)\bigr) 
                                                  \ , \nonumber\\ 
  a(u) &=& \frac{(1-\ru)^3(3+\ru)}{(1+\ru)^3(3-\ru)} \ .
\end{eqnarray} 
In Ref.~\refcite{twoloopsunrise} we found also that 
\begin{equation}\label{clausen}
\int_0^1 du\;J_1(u)=\Cl_2 (\pi/3)\ .
\end{equation}
We consider here the integrals of products of $J_i$: we have checked up to
2400-digits of precision the following equalities:
\begin{align}\label{prodotto01}
\int_0^1 du\;J_1^2(u)&=\dfrac{1}{8}A_4\ ,\\
\int_0^1 du\;J_2^2(u)&=\dfrac{3}{4}A_4\ ,\\
\label{prodotto013}
\int_0^1 du\;J_1(u)J_2(u)&=\dfrac{1}{4}B_4\ ,
\end{align}
and 
\begin{align}\label{prodotto19}
\int_1^9 du\;J_1^2(u)&=\dfrac{3}{8}A_4\ ,\\
\int_1^9 du\;J_2^2(u)&=\dfrac{9}{8}A_4\ ,\\
\label{prodotto193}
\int_1^9 du\;J_1(u)J_2(u)&=\dfrac{1}{2}B_4\ .
\end{align}
Therefore we have identified some one-dimensional integral representations of
the numerical constants $A_4$ and $B_4$.

\subsection{The key observation}
\eqrefb{prodotto01}{prodotto193} are not satisfactory  
elementary definitions of $A_4$ and $B_4$, because of the complexity 
of the arguments of the $K$ function in \eqref{exp01}.
Our aim is to find out integral representations of $A_4$ and
$B_4$ as simple as \eqref{I2001} and \eqref{I1101}.
We make a comparison between the integrals of products of $K$ of the family \itref{family} 
and the integrals of products of $J_i$.
\eqref{clausen}  corresponds to 
\begin{equation}\label{I101} 
\int^1_0 dt\;\dfrac{K(t)}{\sqrt{1-t}}=\dfrac{\pi^2}{2} 
\end{equation} 
We try to modify the integrand of  \itref{I101} so that the result
contains the constant $\Cl_2(\pi/3)$.
Our many year experience with the analytical calculation of 3-loop $g$-$2$
suggests that such a constant is usually associated with integrals containing 
the polynomial $1+t+t^2$ in the denominator. This is a factor of $1-t^3$.
Therefore we try to consider a ``cubic'' modification
of the usual elliptic integral
$K(m)$.
One fruitful choice is 
\begin{equation}
K_c(m)=\int_0^1 \dfrac{dt}{\sqrt[3]{(1-t^3)(1-mt^3)^2}}\ , \qquad
E_c(m)=\int_0^1 \dfrac{dt\;\sqrt[3]{1-mt^3}}{\sqrt[3]{1-t^3}}\ , 
\end{equation}
or, equivalently, expressing them in terms of the hypergeometric function
\begin{equation}\label{cubicke}
K_c(m)=\dfrac{2\pi}{\sqrt{27}} \;\Fcub{m}\ , \qquad
E_c(m)=\dfrac{2\pi}{\sqrt{27}} \;\Ecub{m}\ .
\end{equation}
Now we calculate the numerical values of the integrals
obtained by replacing $K$ with $K_c$ in \eqref{family},
and we look for relations with the 
constants $A_4$ and $B_4$.
Luckily, we find
\begin{align}
\label{ab4cubic}
A_4&=\dfrac{9}{5}\int_0^1 dx \dfrac{K_c(x)K_c(1-x)}{\sqrt{1-x}} \ ,\\
\label{ab4cubic2}
B_4&=\dfrac{3\sqrt{3}}{4}\int_0^1 dx \dfrac{K_c^2(x)}{\sqrt{1-x}}  \ .
\end{align}
We stress the tremendous simplification
obtained by going from the usual description with elliptic integrals
\itrefb{exp01}{prodotto013} to the ``cubic'' version
\itrefb{cubicke}{ab4cubic2}.

\section{Four-loop results}\label{risultati}
For the sake of brevity we define the following constants
\begin{align}\label{cdef}
C&=\int_0^1 dx \dfrac{K_c^2(x)}{\sqrt{1-x}} =
7.396\;099\;534\;768\;919\;553\;449\;114\;417\;961\;526\;519\;642{\ldots}\ ,\\
\cal D&=\int_0^1 dx \dfrac{K_c(x)K_c(1-x)}{\sqrt{1-x}} =\nonumber\\
&\qquad\qquad\qquad\qquad
4.860\;756\;383\;778\;595\;063\;430\;474\;772\;965\;586\;029\;529{\ldots}\ ,\\
E&=\int_0^1 dx \dfrac{E_c^2(x)}{\sqrt{1-x}} = 
2.376\;887\;326\;184\;666\;003\;152\;855\;958\;761\;330\;926\;023\;{\ldots} \ .
\end{align}
Now we look for relations between the numerical values of the above constants
$C$, $\cal D$ and $E$ 
and the numerical values of $S_4(D=2)$   
\eqref{S4_due} and $S_4(D=4-2\e)$
\eqref{S4_quattro}.
We find that
\begin{equation}\label{S4_due_C}
S_4(-1,1,1,1,1,1,D=2)=\pi \sqrt{3} C\ ,
\end{equation}
or, alternatively,
\begin{equation}
S_4(-1,1,1,1,1,1,D=2)=\dfrac{4}{3}\pi B_4\ ;
\end{equation}
we note the appearance of the factor $4\pi/3$, similar to the appearance
of $4\pi/\sqrt{15}$ in \eqref{S3_due}.
By using the integer-relation search program PSLQ\cite{PSLQ} we have been
able to fit the numerical result of \eqref{S4_quattro}
with the analytical expression
\begin{equation}\label{S4_quattro_C}
\begin{split}
S_4&(-1,1,1,1,1,1,D=4-2\e)=
-\frac{5}{2\e^4} -\frac{45}{4\e^3} -\frac{4255}{144\e^2}
-\frac{106147}{1728\e}  +c_0 
+O(\e)\ , \\
c_0&=
 \frac{\pi\sqrt{3}}{240}\left( 297 C - 1477 E \right) 
- \frac{2320981}{20736}\ . 
\end{split}
\end{equation}
The equalities \eqref{S4_due_C} and \eqref{S4_quattro_C} are the main result
of this paper; they have been checked up to 2400 digits of precision.
Note that constant $\cal D$ does not appear in \eqref{S4_due_C} and \eqref{S4_quattro_C}.
\section{Four-dimensional lattice integrals}\label{quattrolattice}

Considering lattice perturbation theory,  at one loop level one finds
these integrals\cite{Capitani}
\begin{equation}\label{defz0}
Z_0 = \int_{-\pi}^{\pi} \dfrac{d^4k}{(2\pi)^4}  \dfrac{1}{4 \sum_{\lambda=1}^4
\sin^2 \left(k_\lambda/2\right)} 
=0.154\;933\;390\;231\;060\;214{\ldots} 
\ ,
\end{equation}
and 
\begin{equation}
Z_1 = \int_{-\pi}^{\pi} \dfrac{d^4k}{(2\pi)^4}  
\dfrac{\sin^2 \left(k_1/2\right)\sin^2 \left(k_2/2\right)}{ \sum_{\lambda=1}^4
\sin^2 \left(k_\lambda/2\right)}
=0.107\;781\;313\;539\;874\;001{\ldots} 
\ ,
\end{equation}
Some years ago, while visiting the Department of Physics of Parma,
York Schr\"oder pointed out to the author that,
whether 3-loop $g$-$2$ was known in analytical form,
no analytical result was known for the ``simple'' lattice 1-loop tadpole $Z_0$.
Puzzled by this fact,
and noting that $Z_0$ can be reduced to a triple elliptic integral,
we have tried to relate the numerical values of $Z_0$ and $Z_1$ 
to the new constants $C$, $\cal D$ and $E$.
Working with only 10-digits precision numbers we have discovered numerically that 
\begin{equation}
\dfrac{S_4(-1,1,1,1,1,1,D=2)}{\pi^4 Z_0}\approx8/3\ .
\end{equation}
That is 
\begin{equation}\label{z0}
Z_0 \pi^3=\dfrac{3\sqrt{3}}{8} C   \ .
\end{equation}
By using again PSLQ, we have also found that
\begin{equation}\label{z1}
Z_1 \pi^3 = - \dfrac{\sqrt{3}}{20} \left( 3C + 7E\right) 
+ \dfrac{\pi^3}{4} 
- \dfrac{\pi}{3}  \ .
\end{equation}
Values of $Z_0$ and $Z_1$ with 400 digits of precision are quoted in
Ref.~\refcite{Capitani}. Very kindly, York Schr\"oder provided us 
16000-digits values. By using these numbers, 
we have checked \eqref{z0} and \eqref{z1} up to 2400 digits of precision, 
the maximum precision of our values of $C$ and $E$.

We note also that the integral \itref{defz0} can be rewritten 
into the so-called Watson integral in 4-dimensions 
(see Refs.~\refcite{Joyce3,glasser})
\begin{equation}
u(4)=\dfrac{4}{(2\pi)^4} 
\int_{-\pi}^{\pi}
\int_{-\pi}^{\pi}
\int_{-\pi}^{\pi}
\int_{-\pi}^{\pi}
\dfrac{dk_1\;dk_2\;dk_3\;dk_4}{4-\cos k_1 -\cos k_2 -\cos k_3 -\cos k_4 } =8\;Z_0  \ .
\end{equation}
From \eqref{z0} one obtains 
\begin{equation}
u(4)\pi^3 = 3\sqrt{3} C  \ .
\end{equation}

\section{Conclusions}\label{conclusions}
In this paper we have at last identified  beyond any reasonable doubt
the analytical constants 
which appear in the simplest non-trivial 4-loop $g$-$2$ master integral.
We have also discovered that the \emph{same} constants appear in 
some 4-dimensional lattice integrals.
Clearly we still do not know a rigorous proof of these relations, but,
once the form of the results is known,
we hope that proofs will be easier to find
(see the very recent papers\cite{bro1,bro2,talkbroadhurst2}).

\subsection[subtitle]{Note on Ref.~\refcite{bro1}}\label{compbro}
While we were completing this paper, kindly David Broadhurst sent us 
a copy of his new paper\cite{bro1}.
In that paper, several elliptic integral evaluations of Bessel moments are
performed.
In particular, a proof of our \eqref{A3} and \eqref{A3_mod} is given,
as well as of \eqref{I2011} and \eqref{I1111}.
Constants analogous to our $A$ and $B$ are found, 
$c_{4,0}=2\pi A$ and $s_{4,0}=B$, as well as $t_{6,1}=A_4/8$ and
$s_{6,1}=S_4(D=2)/16$. In addition, several relations between Bessel moments
are found, and some evaluations of double elliptic integrals are done.

\section{Acknowledgements}
We thanks Enrico Onofri and Giuseppe Amoretti for helping us 
to retrieve the unpublished 3-loop results 
\eqrefb{quattrogamma}{S3_quattro_res} from a remote off-line computer,
on occasion of the talk \cite{talkbroadhurst}.

\end{document}